\documentclass[aps,prd,preprintnumbers,superscriptaddress,nofootinbib,showpacs,twocolumn]{revtex4}%
\usepackage[dvips]{graphicx}
\usepackage{bm,latexsym,amsmath,amssymb,amsfonts,mathrsfs}
\usepackage{color}
\input{colordvi.tex}
\usepackage[dvipdfm]{hyperref}
\usepackage{url}
\hypersetup{
    colorlinks=true,
    citecolor=cyan,
}
\newcommand*{\D}{{\rm d}}

\begin{document}

\title{Suppressing the primordial tensor amplitude without changing the scalar sector
in quadratic curvature gravity}

\author{Kohji~Yajima}
\email[Email: ]{kohji"at"rikkyo.ac.jp}
\affiliation{Department of Physics, Rikkyo University, Toshima, Tokyo 175-8501, Japan
}

\author{Tsutomu~Kobayashi}
\email[Email: ]{tsutomu"at"rikkyo.ac.jp}
\affiliation{Department of Physics, Rikkyo University, Toshima, Tokyo 175-8501, Japan
}

\begin{abstract}
We address the question of how one can modify the inflationary tensor spectrum
without changing at all the successful predictions on the curvature perturbation.
We show that this is indeed possible, and
determine the two quadratic curvature corrections
that are free from instabilities and
affect only the tensor sector at the level of linear cosmological perturbations.
Both of the two corrections can reduce the tensor amplitude, though one of them
generates large non-Gaussianity of the curvature perturbation.
It turns out that the other one corresponds to so-called Lorentz-violating Weyl gravity.
In this latter case one can obtain as small as 65\%
of the standard tensor amplitude. Utilizing this effect we demonstrate that
even power-law inflation can be within the 2$\sigma$ contour of the Planck results.
\end{abstract}

\pacs{98.80.Cq}
\preprint{RUP-15-17}
\maketitle

\section{Introduction}

Inflation~\cite{inflation,Yokoyama:2014nga} plays a crucial role in cosmology of the very early Universe.
In particular, single-field slow-roll models of inflation generically
produce nearly scale-invariant, adiabatic, and Gaussian curvature perturbations
as the seeds for cosmic structure~\cite{fluctuation}.
The theoretical prediction
matches observational results e.g. of the Planck experiments fairly well~\cite{Ade:2013ktc,Ade:2013uln,Adam:2015rua,PlanckCollaboration2015b}.
During inflation primordial tensor modes (gravitational waves) are generated as well.
The tensor amplitude is conventionally parametrized by the tensor-to-scalar ratio, $r$,
and the Planck constraint on $r$ is given by $r<0.10$ (95\%
C.L.)~\cite{PlanckCollaboration2015b}.
Some of the single-field slow-roll models predict larger tensor modes,
and hence have been excluded by current observations.
One would then ask whether one can reduce the tensor amplitude somehow to save such models.
This is the question which we discuss in this paper.

General relativity is an underlying assumption of standard inflation models,
and nonstandard dynamics of the tensor modes can be obtained
by modifying this gravitational sector.
In doing so, one generically expects that
the behavior of the scalar perturbations is also modified.
This is however what we want to avoid, because
the standard inflationary predictions on the scalar perturbations are so successful.
In this paper,
we therefore explore the possibility of
modifying {\em only} the tensor modes and try to retain the same structure
of the scalar sector as in general relativity,
in order not to spoil the remarkable agreement between
the standard theoretical predictions of the scalar perturbations and observations.

It is natural to consider quadratic curvature terms in the action beyond general relativity
since such corrections are expected to arise as signatures of new physics at high energies.
Below we look for quadratic curvature terms that modify only the tensor sector
of cosmological perturbations without introducing any pathologies such as ghost instabilities.
It turns out that there are two independent combinations of the curvature tensors
fulfilling the above requirements.
Both combinations do not change the quadratic action for the scalar perturbations,
and one of them has no impact on the cubic action as well.
The resultant quadratic curvature terms are not of the form
${\cal R}^2$, ${\cal R}_{\mu\nu}{\cal R}^{\mu\nu}$, etc. which are familiar in the literature~\cite{Stelle:1976gc,Stelle1978},
but they have nontrivial coupling to the derivative of the inflaton field.
We study in detail how the tensor amplitude and tilt are modified,
and discuss the implications for observations.

The organization of this paper is as follows.
We determine the two possible quadratic curvature terms
satisfying our requirement in the next section.
In Sec.~III we evaluate the modified amplitude and tilt of the
primordial tensor modes, and then present the implications for observations.
We draw our conclusions in Sec.~IV.

\section{Construction of the Lagrangian}

The theory we consider is described by the action
\begin{eqnarray}
S=S_{\rm EH}+S_\phi +S_{\rm higher},
\end{eqnarray}
where $S_{\rm EH}$ is the Einstein-Hilbert term,
\begin{eqnarray}
S_{\rm EH}=\frac{1}{2\kappa}\int\D^4x\sqrt{-g}{\cal R},
\end{eqnarray}
$S_\phi$ is the action of the inflaton field,
\begin{eqnarray}
S_\phi =\int\D^4x\sqrt{-g}P(\phi, \partial^\mu \phi\partial_\mu\phi),
\end{eqnarray}
and $S_{\rm higher}$ represents higher curvature corrections,
\begin{eqnarray}
S_{\rm higher}=\frac{1}{\kappa} \int\D^4x\sqrt{-g}\left(
\frac{1}{M^2}{\cal R}_{\mu\nu\rho\sigma}{\cal R}^{\mu\nu\rho\sigma}+\cdots \right).\label{high-ex}
\end{eqnarray}
The simplest Lagrangian for the inflaton field would be of the form
$P=-\partial^\mu\phi\partial_\mu\phi/2-V(\phi)$,
but here we do not need to specify the concrete form of $P$.

It is known that
typical higher curvature terms like the one presented in Eq.~(\ref{high-ex})
give rise to new propagating degrees of freedom
which are plagued by (ghost) instabilities~\cite{Ostrogradski}.
In this paper, we carefully construct higher curvature terms
so that the resultant theory is free from such dangerous degrees of freedom.
Among such healthy theories we are interested in those in which
the dynamics of tensor perturbations is modified
while the scalar sector of cosmological perturbations is left unchanged.
To find the higher curvature terms having those properties,
we have to go beyond the familiar curvature invariants such as
${\cal R}_{\mu\nu\rho\sigma}{\cal R}^{\mu\nu\rho\sigma}$ and ${\cal R}_{\mu\nu}{\cal R}^{\mu\nu}$,
and consider the terms obtained by contracting with
the unit normal to constant $\phi$ hypersurfaces,
\begin{eqnarray}
u_\mu:=-\frac{\partial_\mu\phi}{\sqrt{-\partial^\nu\phi\partial_\nu\phi}},
\end{eqnarray}
and the induced metric,
\begin{eqnarray}
\gamma_{\mu\nu} = g_{\mu\nu}+u_\mu u_\nu,
\end{eqnarray}
e.g.,
${\cal R}_{\mu\nu\rho\sigma}{\cal R}_{\mu'\nu'\rho'\sigma'}\gamma^{\mu\mu'}\gamma^{\nu\nu'}\gamma^{\rho\rho'}u^\sigma u^{\sigma'}$.
This possibility was demonstrated
in the context of Weyl gravity in Ref.~\cite{Deruelle:2012xv}.

Focusing on quadratic curvature corrections,
we are going to identify the terms in the Lagrangian fulfilling the above requirements
in the following way. The basic idea here is along the same line as taken in Refs.~\cite{GLPV,Gao}.
We start by performing the Arnowitt-Deser-Misner (ADM) decomposition,
taking constant $\phi$ hypersurfaces as constant time hypersurfaces,
as the dynamics of cosmological perturbations is more transparent
in the ADM language.
The metric is written in terms of the ADM variables as
\begin{eqnarray}
\D s^2=-N^2\D t^2+\gamma_{ij}\left(\D x^i+N^i\D t \right)\left(\D x^j+N^j\D t \right).
\end{eqnarray}
The possible quadratic curvature terms in the Lagrangian are exhaustively
written in terms of the three-dimensional geometric quantities as
\begin{align}
\sqrt{\gamma}N &\;\times\; \left\{
K^4,\; K_{ij}K^{ij}K^2,\;\cdots,
\;
R^2,\;R_{ij}R^{ij},\right.
\notag \\
& \left.
K^2R,\; KK^{ij}R_{ij},\;\cdots,
\;
D_iK_{jk}D^iK^{jk},\;\cdots \right\} ,\label{termlist}
\end{align}
where $K_{ij}$ and $R_{ij}$ are
the extrinsic and intrinsic curvature tensors of
the constant $\phi$ hypersurfaces, respectively,
$D_i$ stands for the covariant derivative with respect to $\gamma_{ij}$,
and ellipses are used to indicate analogous terms whose indices are contracted in different ways. 
We discard from the above candidates
the terms containing time derivatives of the extrinsic curvature,
because higher time derivatives of the metric imply the appearance of
additional propagating degrees of freedom other than $\phi$
and two tensor modes, signaling instabilities.\footnote{If
one has only $\dot K$ the theory is not necessarily unstable,
as is illustrated by the example of the ${\cal R}^2$ model.
This however adds an extra scalar degree of freedom
modifying the scalar sector of cosmological perturbations.
For this reason we avoid any time derivatives of $K_{ij}$.}

Let us consider cosmological perturbations,
\begin{align}
N=1+\delta N,\quad N_i=\partial_i\chi +\chi_i,\quad
\gamma_{ij}=a^2e^{2\zeta}\left(e^h \right)_{ij},
\end{align}
where $\zeta$ is the curvature perturbation on the uniform $\phi$ hypersurfaces,
$h_{ij}$ is the transverse and traceless tensor perturbation, and $\chi_i$
is the transverse vector perturbation.
Let us concentrate on the scalar sector for the moment.
To first order in perturbations,
the extrinsic curvature is given by
\begin{eqnarray}
K_i^{\;j} =H\delta_i^{\;j}+\frac{1}{3}\delta K\delta_i^{\;j} +
\delta \widetilde K_i^{\;j},
\end{eqnarray}
with
\begin{align}
\delta K =&\; -3H\delta N+3\dot\zeta-\frac{1}{a^{2}} \partial^2\chi,
\\
\delta\widetilde K_i^{\;j}=&\;-\frac{1}{a^{2}}
\left(\partial_i\partial^j-\frac{1}{3}\delta_i^{\;j}\partial^2\right)\chi,
\end{align}
and the intrinsic curvature is
\begin{eqnarray}
\delta R_i^{\;j}=-\frac{1}{a^2}\left(\partial_i\partial^j+\delta_i^{\;j}\partial^2 \right)\zeta.
\end{eqnarray}
The perturbation of the extrinsic curvature tensor has been decomposed into
its trace and traceless parts.
Using those quantities as building blocks,
one can construct
the following two combinations
of the form listed in Eq.~(\ref{termlist})
for which the scalar-type variables are canceled out
after integration by parts,
\begin{eqnarray}
2\partial_i\delta \widetilde K_{jk}\partial^i\delta\widetilde K^{jk}
-3\partial_i\delta\widetilde K^{ik}\partial^j\delta \widetilde K_{jk},
\label{quad-KK}
\end{eqnarray}
and
\begin{eqnarray}
\delta R_{ij}\delta R^{ij}-\frac{3}{8}\delta R^2,
\end{eqnarray}
at quadratic order in perturbations.
No other combinations can be found with vanishing scalar-type variables.
Now including vector and tensor perturbations we have
\begin{align}
&
2\partial_i\delta \widetilde K_{jk}\partial^i\delta\widetilde K^{jk}
-3\partial_i\delta\widetilde K^{ik}\partial^j\delta \widetilde K_{jk}
\notag \\
&\qquad\qquad\qquad
=\frac{1}{2a^2}\left(\partial_i\dot h_{jk}\right)^2+\frac{1}{4a^6}\left(\partial^2\chi_i \right)^2,
\\
&
\delta R_{ij}\delta R^{ij}-\frac{3}{8}\delta R^2 = \frac{1}{4a^4}\left(\partial^2h_{ij}\right)^2.
\label{RR-tensor}
\end{align}
Both of the two possible quadratic terms for $h_{ij}$ with four derivatives are obtained,
while we successfully exclude $\ddot h_{ij}^2$ which would cause Ostrogradski ghosts.
Since there is no kinetic term for $\chi_i$ here and in $S_{\rm EH}$,
the vector perturbation is not dynamical.
We therefore ignore the vector sector in this paper.

Having thus written the quadratic Lagrangian for perturbations
in terms of the geometric quantities, it is almost straightforward
to determine the full nonlinear Lagrangian in the ADM form as
\begin{align}
{\cal L}_1' =&\;
\frac{\sqrt{\gamma}N}{M^2}
\left(
2D_i \widetilde K_{jk}D^i\widetilde K^{jk}
-3D_i\widetilde K^{ik}D^j \widetilde K_{jk}
\right),
\\
{\cal L}_2 =&\; 
\frac{\sqrt{\gamma}N }{M^2}\left(
R_{ij} R^{ij}-\frac{3}{8} R^2
\right),
\end{align}
where $\widetilde K_{ij}:=K_{ij}-(1/3)K\gamma_{ij}$
is the traceless part of the extrinsic curvature.
Note that ${\cal L}_1'$ is one of the candidates;
in fact, we have different choices
that reduce to Eq.~(\ref{quad-KK}) after integration by parts
at the level of the quadratic Lagrangian for perturbations.
Among them we adopt
\begin{align}
{\cal L}_1 =\frac{\sqrt{\gamma}N}{M^2}
\bigl(&
2D_i \widetilde K_{jk}D^i\widetilde K^{jk}
-D_i\widetilde K^{ik}D^j \widetilde K_{jk}
\notag\\ &
-2D_i\widetilde K_{jk}D^j \widetilde K^{ik}
\bigr)\label{L1true}
\end{align}
rather than ${\cal L}_1'$. What is particular to ${\cal L}_1$
is that it can be written as a square of some tensor as
\begin{eqnarray}
{\cal L}_1=\frac{\sqrt{\gamma}N}{M^2}W_{ijk}W^{ijk},
\end{eqnarray}
where
\begin{eqnarray}
W_{ijk}= 2D_{[i}\widetilde K_{j]k}+D_l\widetilde K^l_{\;[i}\gamma_{j]k}.
\end{eqnarray}
It is clear that the scalar perturbations do not participate in $W_{ijk}$
at first order.
This means that ${\cal L}_1$ does not modify
the scalar sector both
at quadratic and {\em cubic} order.
In other words, the prediction for non-Gaussianity of the curvature perturbation,
as well as that for the power spectrum, remains the same
in the presence of ${\cal L}_1$.
This is however not the case for ${\cal L}_1'$ and ${\cal L}_2$.

The covariant form of the Lagrangian can be recovered
by writing the extrinsic curvature as $K_{\mu\nu}=\gamma_\mu^{\;\rho}\gamma_\nu^{\;\sigma}\nabla_\rho u_\sigma$
and making use of the Gauss-Codazzi relations:
\begin{eqnarray}
\gamma^{\;\mu}_{\alpha}\gamma^{\;\nu}_{\beta}\gamma^{\;\gamma}_{\rho}\gamma^{\;\sigma}_{\delta}{\cal R}^{\rho}_{\;\sigma \mu \nu} &=& R^{\gamma}_{\;\delta \alpha \beta} + K^{\;\gamma}_{\alpha} K_{\delta \beta} - K^{\;\gamma}_{\beta} K_{\alpha \delta} ,\;\;\;\\
u^\mu\gamma_\alpha^{\;\nu}\gamma_\beta^{\;\rho}\gamma_\gamma^{\;\sigma}
{\cal R}_{\mu\nu\rho\sigma} &=& D_{\gamma} K_{\alpha\beta} - D_{\beta}K_{\alpha\gamma}.
\end{eqnarray}
We find
\begin{eqnarray}
{\cal L}_1 = \frac{\sqrt{-g}}{M^2}
C_{\mu\nu\rho\sigma}C_{\mu'\nu'\rho'\sigma'}\gamma^{\mu\mu'}\gamma^{\nu\nu'}\gamma^{\rho\rho'}u^\sigma u^{\sigma'},
\end{eqnarray}
where $C_{\mu\nu\rho\sigma}$ is the Weyl tensor.
Thus, it turns out that
${\cal L}_1$ reproduces the theory studied in Ref.~\cite{Deruelle:2012xv}.
One can repeat the same procedure also for ${\cal L}_2$ to derive its covariant form.
However, the covariant expression for ${\cal L}_2$ is messy and not so illuminating,
so that in this case it is better to work in the simpler ADM form.
It is worth noting that the covariant expression for ${\cal L}_2$ is also constructed
by contracting the Riemann curvature tensor with $u^\mu$ and hence it is Lorentz violating
in the same sense that ${\cal L}_1$ is.

In ${\cal L}_1$ and ${\cal L}_2$ in the ADM form one may consider time-dependent $M$.
This translates to the $\phi$-dependent coupling in the covariant language.
However, since $\phi$ is slowly rolling, it is reasonable to assume that
$M$ is only weakly time dependent. For simplicity we treat
$M$ as constant in the following.

\section{The tensor amplitude}

In the previous section we have identified the
two possible quadratic curvature terms
that make no contribution to the scalar sector of cosmological perturbations
at least at linear order.
Let us now investigate how
the amplitude of primordial tensor modes is modified due to those terms.
For clarity we study each term separately below.
Actually, we will find that a sizable modification
from ${\cal L}_2$ is prohibited because ${\cal L}_2$ also produces large non-Gaussianity.

\subsection{${\cal L}_1$}

First we consider
\begin{eqnarray}
S_{\rm higher}=\frac{1}{\kappa}\int\D^4x{\cal L}_1.
\end{eqnarray}
The quadratic action for the tensor perturbations is given by~\cite{Deruelle:2012xv}
\begin{align}
S=\frac{1}{8\kappa}\int\D t\D^3x\,a^3\left[
\dot h_{ij}^2-\frac{1}{a^2}(\partial_kh_{ij})^2
+\frac{4}{M^2a^2}(\partial_k\dot h_{ij})^2
\right].
\end{align}
Each Fourier mode of two polarization states, $h_k^\lambda(t)$ ($\lambda=+,\times$),
obeys a second-order evolution equation.
We use the canonically normalized variable
\begin{eqnarray}
f_k^\lambda(t) = \left(\frac{1}{4\kappa}\right)^{1/2}a^{3/2}\left(1+\frac{4k^2}{M^2a^2}\right)^{1/2} h_k^\lambda,
\end{eqnarray}
and omit $\lambda$ when unnecessary.
We then have
\begin{eqnarray}
\ddot f_k + \omega^2_k(t) f_k = 0,\label{evolution-f}
\end{eqnarray}
where
\begin{align}
\omega_k^2:= &\; -\frac{1}{4}\left(H^2+2\dot H \right)
+\frac{k^2/a^2-2H^2-\dot H}{1+4k^2/M^2a^2}
\notag \\
&\; -\frac{4H^2k^2/M^2a^2}{(1+4k^2/M^2a^2)^2}.\label{omega-eff}
\end{align}

We use the WKB solution,
\begin{eqnarray}
f_k\simeq \frac{1}{\sqrt{2\omega_k}}\exp\left[-\mathrm{i}\int^t\omega_k(t')\D t'\right],
\label{WKB-sol}
\end{eqnarray}
for the short wavelength modes with $k^2/a^2\gg H^2, M^2$.
In the short wavelength limit, Eq.~(\ref{omega-eff}) reduces to
\begin{eqnarray}
\omega_k^2 \simeq \frac{M^2}{4}-\frac{1}{4}\left(H^2+2\dot H \right).\label{omega-approx}
\end{eqnarray}
From this we see that during inflation the tensor perturbations are stable
if
\begin{eqnarray}
H< M.\label{stability}
\end{eqnarray}
We assume that the background evolution,
which is controlled by the inflaton sector $S_\phi$,
satisfies the condition~(\ref{stability}).
Since Eq.~(\ref{omega-approx}) gives the estimate
\begin{eqnarray}
\frac{\dot\omega_k}{\omega_k^2} \sim \frac{\epsilon H^3}{(M^2-H^2)^{3/2}},
\end{eqnarray}
where $\epsilon:=-\dot H/H^2\ll1$ is the slow-roll parameter,
the WKB approximation is justified as long as
$H$ is not too close to $M$.

In the long wavelength limit, $k^2/a^2\ll H^2, M^2$, we have
\begin{eqnarray}
\omega_k^2\simeq -\frac{9}{4}H^2-\frac{3}{2}\dot H=-\frac{(a^{3/2})\,\ddot{}}{a^{3/2}},
\end{eqnarray}
so that the standard result is recovered on superhorizon scales, $h_k\simeq$ const.

Let us compute the power spectrum of
the tensor modes,
\begin{eqnarray}
{\cal P}_T(k)=\frac{k^2}{\pi^2}\left|h_k \right|^2.
\end{eqnarray}
In general, Eq.~(\ref{evolution-f}) cannot be solved analytically,
and hence one needs numerical calculations to evaluate the power spectrum.
However, in the special case of the exact de Sitter background
one can solve Eq.~(\ref{evolution-f}) analytically
using the hypergeometric functions.
This was done in Ref.~\cite{Deruelle:2012xv}, and here
we quote their final result:
\begin{eqnarray}
{\cal P}_T=\frac{2\kappa H^2}{\pi^2}\Xi_1(H/M),\label{dS-formula}
\end{eqnarray}
where
\begin{eqnarray}
\Xi_1(x):=\frac{\cosh(\pi\nu/2)\coth(\pi\nu/2)|\Gamma(-1/4+\mathrm{i}\nu/4)|^4}{128\pi^2x^3},
\end{eqnarray}
with $\nu:=\sqrt{x^{-2}-1}$.
Based on this, one may expect that in the case where $H$ is varying
the power spectrum is given by
evaluating the de Sitter result~(\ref{dS-formula}) at horizon crossing,
\begin{eqnarray}
{\cal P}_T(k)=\left. \frac{2\kappa H^2}{\pi^2}\Xi_1(H/M)\right|_{k=aH},
\label{PT-at-hc}
\end{eqnarray}
as is commonly done in general relativity.

\begin{figure}[tb]
  \begin{center}
    \includegraphics[keepaspectratio=true,height=52mm]{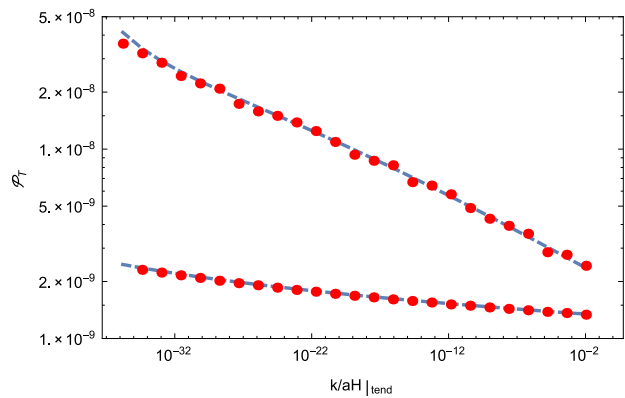}
  \end{center}
  \caption{
The power spectrum ${\cal P}_T$ of tensor modes from power-law inflation with the correction from ${\cal L}_1$.
The upper line and points are for $a\propto t^{50}$ and the lower for $a\propto t^{400}$, respectively.
  Red points represent the numerical results, while the dashed line indicates the analytic estimate~(\ref{PT-at-hc}).
  The parameters for the upper line and points are given by $\sqrt{\kappa}H|_{t_{\rm end}}=10^{-4}$ and $H/M|_{t_{\rm end}}=0.16$,
  where $t_{\rm end}$ is the time at the end of inflation.
  The parameters for the lower ones are given by $\sqrt{\kappa}H|_{t_{\rm end}}=10^{-4}$ and $H/M|_{t_{\rm end}}=0.74$.
  }%
  \label{fig:analyticsp.eps}
\end{figure}

We numerically solved Eq.~(\ref{evolution-f}) in the case of power-law inflation, $a\propto t^p$,
using the initial condition~(\ref{WKB-sol}),
and verified that Eq.~(\ref{PT-at-hc}) reproduces
the numerical results very accurately, as shown in Fig.~\ref{fig:analyticsp.eps}.
We are thus allowed to use the formula~(\ref{PT-at-hc})
for slow-roll inflation.

\begin{figure}[tb]
  \begin{center}
    \includegraphics[keepaspectratio=true,height=52mm]{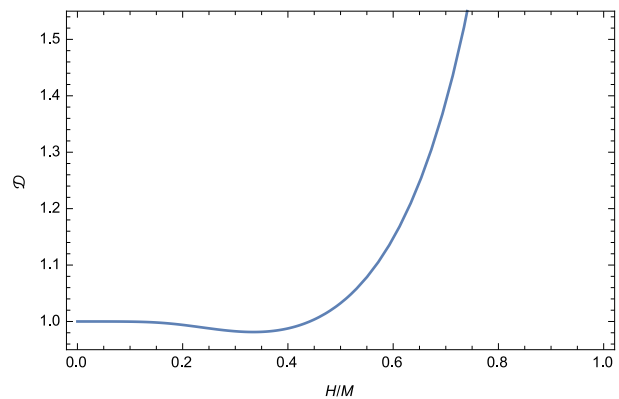}
  \end{center}
  \caption{${\cal D}$ as a function of $H/M$.
  }%
  \label{fig:consis_weyl.eps}
\end{figure}

The behavior of the function $\Xi_1(x)$ is as follows:
$\Xi_1\to 1$ as $x\to 0$, and $\Xi_1<1$ for $0<x\lesssim 0.95$.
Thus, the tensor amplitude is suppressed for $H\lesssim 0.95 M$.
The minimum of $\Xi_1$ is given by $\Xi_1\simeq 0.65$, which occurs at $x\simeq 0.74$, and
$\Xi_1$ diverges as $x\to 1$.

Since the power spectrum of the curvature perturbation remains unchanged
in our theory, the tensor-to-scalar ratio is given by
\begin{eqnarray}
r=16\epsilon \Xi_1. \label{eq-r}
\end{eqnarray}
The tensor tilt, $n_T:=\D\ln{\cal P}_T/\D\ln k$, is evaluated as
\begin{eqnarray}
n_T=-\frac{2\epsilon}{1-\epsilon}
\left.\left[1+\frac{1}{2}\frac{\D\ln\Xi_1}{\D\ln(H/M)} \right]\right|_{k=aH}.\label{eq-nt}
\end{eqnarray}
We see that $\D\ln \Xi_1/\D\ln x < 0$ for $0<x\lesssim 0.74$
and its minimum value is given by $\D\ln \Xi_1/\D\ln x\simeq -0.46$ at $x\simeq 0.53$.
This shows that the tensor spectrum is always red.

From Eqs.~(\ref{eq-r})--(\ref{eq-nt}) it is clear that
the consistency relation~\cite{Lidsey} is violated.
The deviation from the standard consistency relation is characterized by
the following function,
\begin{eqnarray}
{\cal D}:=\left.\frac{1+(1/2)\D \ln\Xi_1/\D \ln x}{\Xi_1} \right|_{x=H/M},
\end{eqnarray}
as $-8n_T/r\simeq {\cal D}|_{k=aH}$.
In Fig.~\ref{fig:consis_weyl.eps}, we plot ${\cal D}$ as a function of $H/M$.
We see that the violation depends on the scale $k$, and Fig.~\ref{fig:consis_weyl.eps}
tells us its scale dependence.

\begin{figure*}[htb]
  \begin{center}
    \includegraphics[keepaspectratio=true,height=90mm]{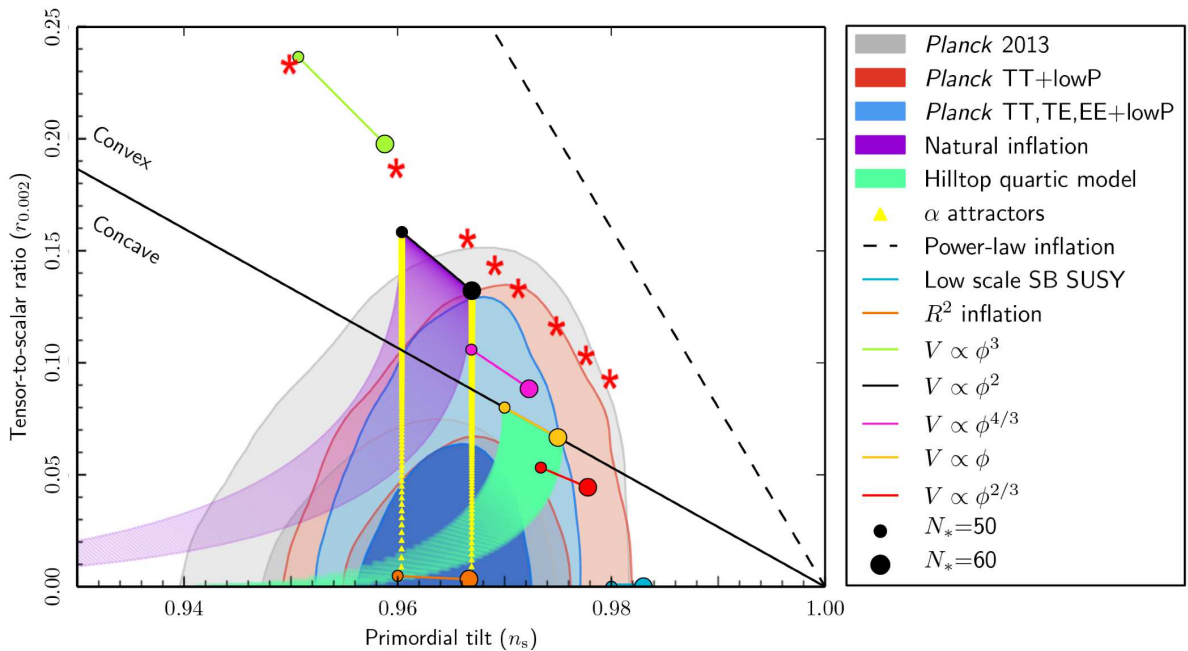}
  \end{center}
  \caption{Reduced tensor amplitude and the Planck results~\cite{PlanckCollaboration2015b} in the $n_s$-$r$ plane.
  Red stars correspond to power-law inflation with the ${\cal L}_1$ correction, assuming the
  maximal suppression, $\Xi_1=0.65$.
  }%
  \label{fig:planck_const6_1.eps}
\end{figure*}

Figure~\ref{fig:planck_const6_1.eps} illustrates the observational implications of the ${\cal L}_1$ correction
by comparing the suppressed tensor amplitude with the Planck results.
The red stars in the figure indicate the case of power-law inflation,
assuming the maximal suppression ($\Xi_1=0.65$) at the observed scale.
Although original power-law inflation (represented by the dashed line)
is ruled out by observations, it can be within the 2$\sigma$ contour with the help of ${\cal L}_1$.
The same applies to other inflation models such as $V\propto \phi^2$.
Those models originally predict large tensor modes,
but the ${\cal L}_1$ correction can bring such models to
the observationally preferred region in the $n_s$-$r$ plane.

\subsection{${\cal L}_2$}

Next let us consider
\begin{eqnarray}
S_{\rm higher}=- \frac{1}{2\kappa}\int\D^4x{\cal L}_2.
\end{eqnarray}
Here we added a minus sign so that the
tensor perturbations are stable at high momenta.
The quadratic action for the tensor perturbations is
\begin{align}
S=\frac{1}{8\kappa}\int\D t\D^3x\,a^3\left[
\dot h_{ij}^2-\frac{1}{a^2}(\partial_kh_{ij})^2
-\frac{1}{M^2a^4}(\partial^2 h_{ij})^2
\right].
\end{align}
Using the conformal time defined by $\D \eta = \D t/a$ and
the canonically normalized variable $v_k^\lambda:=(4\kappa)^{-1/2}a h_k^\lambda$ in the Fourier space,
we obtain
\begin{eqnarray}
\frac{\D^2v_k}{\D\eta^2}+\omega^2_k(\eta)v_k = 0,\label{veom}
\end{eqnarray}
where
\begin{eqnarray}
\omega_k^2:=k^2+\frac{k^4}{M^2a^2}-\frac{1}{a}\frac{\D^2a}{\D\eta^2}.
\end{eqnarray}
This modified dispersion relation has been studied in detail
in the literature~\cite{Martin:2002kt,Ashoorioon:2011eg}.

The WKB solution
\begin{eqnarray}
v_k\simeq \frac{1}{\sqrt{2\omega_k}}\exp\left[-\mathrm{i}\int^\eta\omega_k(\eta')\D\eta'\right],
\label{vWKB}
\end{eqnarray}
may be used for the short wavelength modes, because $\omega_k^{-2}\D\omega_k/\D\eta \ll 1$
is always satisfied at large $k$.

Only in the case of exact de Sitter inflation for which
the scale factor is given by $a=1/H(-\eta)$,
Eq.~(\ref{veom}) can be solved analytically.
The solution that matches Eq.~(\ref{vWKB}) at large $k$ is obtained
in terms of the Whittaker function as~\cite{Ashoorioon:2011eg,Kobayashi:2015gga}
\begin{eqnarray}
v_k=\frac{e^{-\pi/8x}W_{\mathrm{i}/4x,3/4}(-\mathrm{i}xk^2\eta^2)}{\left(-2xk^2\eta\right)^{1/2}},
\end{eqnarray}
where $x=H/M$. This yields the power spectrum
\begin{eqnarray}
{\cal P}_T=\frac{2\kappa H^2}{\pi^2}\Xi_2(H/M),\label{P2dS}
\end{eqnarray}
where
\begin{eqnarray}
\Xi_2(x):=\frac{\pi}{4}\left[e^{\pi/(4x)}x^{3/2}\left|\Gamma(5/4+\mathrm{i}/(4x))\right|^2 \right]^{-1}.
\end{eqnarray}
In the case of slow-roll inflation, one may evaluate the de Sitter result~(\ref{P2dS})
at horizon crossing, $k=aH$. This can also be justified by a numerical calculation.

One sees that $\Xi_2$ is a monotonically decreasing function
and $\Xi_2\to 1$ as $x\to 0$. Therefore, also in this case
the tensor amplitude is suppressed relative to the standard result.
Since $\Xi_2\propto x^{-3/2}$ for $x\gg 1$, the tensor amplitude could potentially be
suppressed to a very small level.
However, as we see below, this possibility is hindered by
the generation of large non-Gaussianity.

In contrast to the case of ${\cal L}_1$, the cubic action for the curvature perturbation
is affected by ${\cal L}_2$. This implies that $M$ must be sufficiently large
in order to avoid large non-Gaussianities in $\zeta$.
Typically, ${\cal L}_2$ contains terms such as
\begin{eqnarray}
{\cal L}_2 \sim \frac{1}{ M^{2}} \zeta (\partial ^{2} \zeta)^{2}.
\end{eqnarray}
The non-Gaussianity generated by this term is estimated to be
\begin{eqnarray}
f_{\rm NL} \sim \frac{H^{2}}{\epsilon M^{2}} \,.
\end{eqnarray}
Requiring that $f_{\rm NL}\lesssim 1$, we have
\begin{equation}
\frac{H}{M} \lesssim \epsilon^{1/2} \ll 1.
\end{equation}
Therefore, in fact the suppression factor $\Xi_2$ cannot be much smaller than 1.
We conclude that the second Lagrangian ${\cal L}_2$ does not provide
an efficient way of suppressing the tensor amplitude.

One may consider a combination of the two Lagrangians, $a{\cal L}_1+b{\cal L}_2$.
Obviously, this does not change the quadratic Lagrangian for the scalar perturbations, and
to avoid large non-Gaussianities we must require $b\ll 1$.
Therefore, to suppress the tensor amplitude most effectively, essentially one can only use ${\cal L}_{1}$.

\section{Conclusions}

In this paper, we have studied inflationary predictions of theories with quadratic curvature corrections.
We began by looking for ghost-free quadratic curvature terms
that retain the same quadratic action for the curvature perturbation as in general relativity
while modifying the dynamics of tensor perturbations.
We have shown that such curvature terms can indeed be constructed, and
determined the two possible combinations (denoted by ${\cal L}_1$ and ${\cal L}_2$).
This was done by using the ADM formalism, and
recast in a covariant form those corrections contain
the curvature tensors contracted with the unit normal $u^\mu$ to
hypersurfaces on which the inflaton is homogeneous. It has turned out that
one of the two terms, ${\cal L}_1$, is in fact identical to the one introduced in
so-called Lorentz-violating Weyl gravity~\cite{Deruelle:2012xv}.
This term does not change the action of the curvature perturbation even at cubic order.
The other term, ${\cal L}_2$, in contrast, modifies the scalar sector at cubic order.

We have investigated the tensor amplitude in the presence of the
quadratic curvature corrections ${\cal L}_1$ and ${\cal L}_2$.
The analytic results were known only for exact de Sitter inflation,
and we have used the de Sitter formulas evaluated at horizon crossing
in the case of slow-roll inflation for which the Hubble parameter is varying.
The validity of the method has been checked by performing numerical calculations.
Both ${\cal L}_1$ and ${\cal L}_2$ reduce the amplitude of
primordial tensor perturbations. However, we have found that
${\cal L}_2$ could generate large non-Gaussianity of the curvature perturbation,
which places a stringent constraint on the amount of the suppression due to ${\cal L}_2$.
Since ${\cal L}_1$ does not change the cubic interaction of the curvature perturbation,
this evades the non-Gaussianity constraint.
The tensor power spectrum can be as small as 65\%
of the standard result due to ${\cal L}_1$, which
brings many inflation models with large tensor modes to the observationally preferred region
in the $n_s$-$r$ plane.
We have seen that the tensor tilt is also modified, though the spectrum can never be blue.

\acknowledgments
The authors thank Yuuiti Sendouda for helpful comments and discussion.
K.Y. was supported in part by Rikkyo University Special Fund for Research.
This work was supported in part by JSPS Grant-in-Aid for Young
Scientists (B) Grant No.~24740161 (T.K.).




\end{document}